\documentclass{emulateapj}

\def\apjl{ApJL }
\def\aj{AJ }
\def\apj{ApJ }
\def\pasp{PASP }

\def\apjs{ApJS }

\def\aap{A\&A }
\def\nat{Nature }
\def\sun{$_\odot$}
\def\earth{$_\oplus$}

\begin{document}

\title{Stellar Parameters for HD 69830, a Nearby Star with Three Neptune Mass Planets and an Asteroid Belt}

\author{Angelle Tanner\altaffilmark{1}, 
Tabetha S. Boyajian\altaffilmark{2},
Kaspar von Braun\altaffilmark{3},
Stephen Kane\altaffilmark{4},
John M. Brewer\altaffilmark{2},
Chris Farrington\altaffilmark{5},
Gerard T. van Belle\altaffilmark{3},
Charles A. Beichman\altaffilmark{6},
Debra Fischer\altaffilmark{2},
Theo A. ten Brummelaar\altaffilmark{5},
Harold A. McAlister\altaffilmark{5},
Gail Schaefer\altaffilmark{5}}

\altaffiltext{1}{Mississippi State University, Department of Physics \& Astronomy, Hilbun Hall, Starkville, MS, 39762, USA}
\altaffiltext{2}{Department of Astronomy, Yale University, New Haven, CT 06511}
\altaffiltext{3}{Lowell Observatory, 1400 W. Mars Hill Road, Flagstaff, AZ, 86001, USA}
\altaffiltext{4}{Department of Physics and Astronomy, San Francisco State University, San Francisco, CA 94132, USA}
\altaffiltext{5}{Center for High Angular Resolution Astronomy and Department of Physics and Astronomy, Georgia State University, PO Box 4106, Atlanta, GA 30302-4106, USA}
\altaffiltext{6}{NASA Exoplanet Science Institute, California Institute of Technology, MC 100-22, Pasadena, CA 91125, USA}

\begin{abstract}

We used the CHARA Array to directly measure the angular diameter of  HD 69830, home to three Neptune mass planets and an asteroid belt. Our measurement of 0.674$\pm$0.014 milli-arcseconds for the limb-darkened angular diameter of this star leads to a physical radius of R$_*$ = 0.9058$\pm$0.0190 R\sun  and luminosity of  L$_*$ = 0.622$\pm$0.014 L\sun  when combined with a fit to the spectral energy distribution of the star. Placing these observed values on an Hertzsprung-Russel (HR) diagram along with stellar evolution isochrones produces an age of $10.6\pm4$ Gyr and mass of 0.863$\pm$0.043  M\sun. We use archival optical echelle spectra of HD 69830 along with an iterative spectral fitting technique to measure the iron abundance 
([Fe/H]=$-$0.04$\pm$0.03), effective temperature (5385$\pm$44 K) and surface gravity ($\log g = 4.49\pm0.06$). We use these new values for the temperature and luminosity to calculate a more precise age of $7.5\pm3$~Gyr.  Applying the values of stellar luminosity and radius to recent models on the optimistic location of the habitable zone produces a range of 0.61-1.44 AU; partially outside the orbit of the furthest known planet (d) around HD 69830. Finally, we estimate the snow line at a distance of 1.95$\pm$0.19 AU, which is outside the orbit of all three planets and its asteroid belt. 

\end{abstract}


\section{Introduction}

HD 69830, a nearby (12.5 pc, van Leeuwen 2007) K0V star, first came to the attention of the community as one of a few stars with a Spitzer MIPS 24 micron excess indicative of a debris disk (Beichman et al. 2005). Spitzer low resolution IRS (6.5-35 micron) data revealed a complex spectrum of dust emission with features attributed to fosterite, sulfides and carbonates (Beichman et al. 2005; Lisse et al. 2006). Models of the spectral energy distribution (SED) of the infrared excess of HD 69830 and mid-infrared VLTI/MIDI interferometry (Smith et al. 2009) have shown that the debris disk is located at a distance of 0.5-1 AU and is, therefore, analogous to the asteroid belt in our own solar system. More in depth Spitzer observations have concluded that the dust emission is most likely due to replenishment from the collisions of asteroids (Beichman et al. 2011).  Soon after the original Spitzer results were published, three Neptune-mass planets were detected around the star through precision radial velocity (RV) measurements (Lovis et al. 2006) making HD 69830 one of the most unique exoplanetary systems known to date with both a prominent asteroid belt and planetary system. 

Based on the formulation of Kasting et al. (1993) the habitable zone around HD 69830, a K0V star, is located at 0.74$-$0.89 AU, outside of the orbit of the outermost planet, HD 69830d (18.4 M\earth at 0.63 AU, Lovis et al. 2006). Dynamical studies of the system have concluded that an Earth-mass planet within this habitable zone would be dynamically stable (Ji et al. 2007; Rugheimer et al. 2007). This theoretical Earth-mass planet could be detectable with the high precision ($\sim$0.1 m/s) RV measurements which are expected to be available in the next few years. The potential for the future discovery of a planet in the classical habitable zone in this system makes the model-independent knowledge of the radius and luminosity of this star a high priority because then we would be able to characterize the extent of the habitable zone based on empirical data. It is also critical to know the age of HD 69830 for the analysis of the dynamical evolution of the planets as well as to further determine the origin of the dust emission. Previous estimates of the age of the star have ranged from 0.6 to 4.7 Gyr depending on the method used for the determination (Beichman et al. 2005). The age of HD 69830 can dramatically affect our interpretation of the origin of the dust since a young system ($<$1 Gyr) could represent a steady state of dust production while an older ($>$1 Gyr) system could imply a short-lived event.

Here, we present the result of our efforts to directly measure the diameter of HD 69830 with the CHARA interferometric telescope array. In Section 2 we discuss the data and methods used to make the measurement, in Section 3 we review how our measurement can be used to determine additional physical parameters of the star when combined with published photometry and in Section 4 we discuss the implications of these measurements and what information they add to our understanding of this planetary system. 

\section{Observations and Analysis}

We observed HD 69830 for three nights, 2012 Feb 3, 2012 Nov 3 and 2012 Nov 4, using the Georgia State University Center for High Angular Resolution Astronomy (CHARA) Array (ten Brummelaar et al. 2005). The measurements were collected with the S1E1 (330 m) and S1W1 (278 m) CHARA baselines in the H-band (1.67 $\micron$). We used the CHARA Classic beam combiner (Sturmann et al. 2003; ten Brummelaar et al. 2005) in single-baseline mode. Table~\ref{obs} summarizes our observations. Each integration lasted around 2.5 minutes with the 1.5 minutes needed for slewing time between HD 69830 and a set of unresolved calibrator stars, also listed in Table~\ref{obs}. During the observations, HD 69830 was at an elevation of 40 to 45 degrees above the horizon. While this limited the amount of time we had to complete the CHARA observations, the low elevation of the star did not prohibit us from establishing fringes on the science target or its calibration stars. 

\begin{figure}[ht]
\plotone{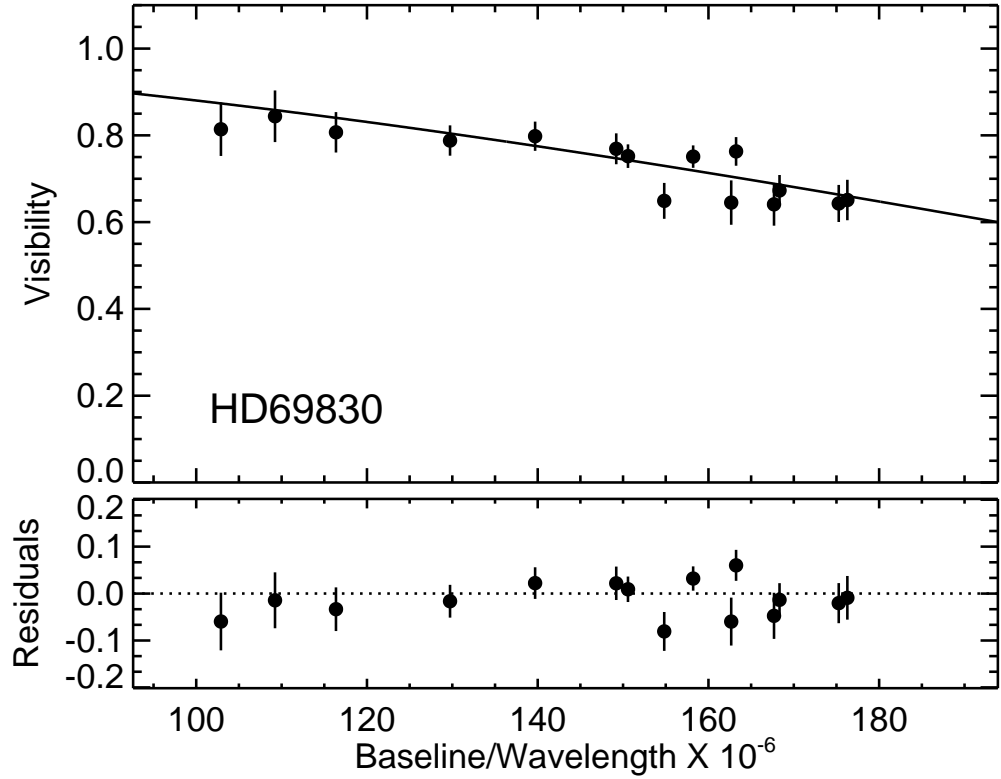}
\figcaption{ Calibrated visibility observations along with the limb-darkened angular fit for HD 69830. For details see $\S$ 2.  \label{uvplot}}
\end{figure}

Our observations employed the common method of bracketing the interferometric observations of the science target with identical observations of calibrator stars which are known to be spatially unresolved at the observing wavelength and baseline. To minimize errors introduced by the calibrator diameters, multiple calibrators are observed with the science star. The surface brightness relations from SearchCal are empirically derived based on data of stars with measured angular diameters.  The the error in the surface brightness angular diameter for the calibrators used is $\sim 7$\%.  The uncertainty in calibrator angular diameter is propagated to the final error in the calibrated visibility of the science target (method described in Boyajian (2009). In our case, we observed HD 71766 ($\theta$ = 0.447$\pm$0.031 milli-arcseconds (mas)), HD 66643  ($\theta$ = 0.439$\pm$0.030 mas) and HD 66242  ($\theta$ = 0.483$\pm$0.033 mas). The calibrator diameters are from the JMMC Stellar Diameters Catalog (Lafrasse et al. 2010), created with the SearchCal tool developed by the JMMC Working Group (Bonneau 2011). These stars were chosen to be point-like and of comparable H-band brightness to HD 69830. Interlacing observations of HD 69830 with these calibrators allowed us to correct for systematics introduced by the instrument and the atmosphere. Figure~\ref{uvplot} shows the visibility measurements as a function of telescope baseline along with the corresponding best fitting limb-darkened model. The coefficients are initially determined from an estimated effective temperature and gravity of the star ($T_{\rm eff}$ = 5400 K and $\log g$=4.5). The coefficients are iterated upon until consistent agreement is found with the final $T_{\rm eff}$ (see Section 3.1). From the CHARA data we get a value of 0.674$\pm$0.014 mas for the limb-darkened angular diameter ($\theta_{\rm LD}$) of HD 69830, using a linear limb-darkening coefficient of $\mu_\lambda$=0.352 (Claret 2000). Our angular diameter measurement of HD~69830 is consistent with the estimate of $\theta_{\rm SB} = 0.65 \pm 0.02$~mas from the $(V-K)$ surface brightness relations in Boyajian et al. (2014). The combination of this angular diameter with a precise Hipparcos parallax, yields a physical stellar radius of R$_*$ = 0.9058$\pm$0.0190 R\sun. All stellar parameters are listed in Table~\ref{props} along with their propagated errors. 

\section{Results}

\subsection{Effective Temperature and Luminosity Determination}

\begin{figure*}
  \begin{center}
    \begin{tabular}{cc}
      \includegraphics[width=9.5cm]{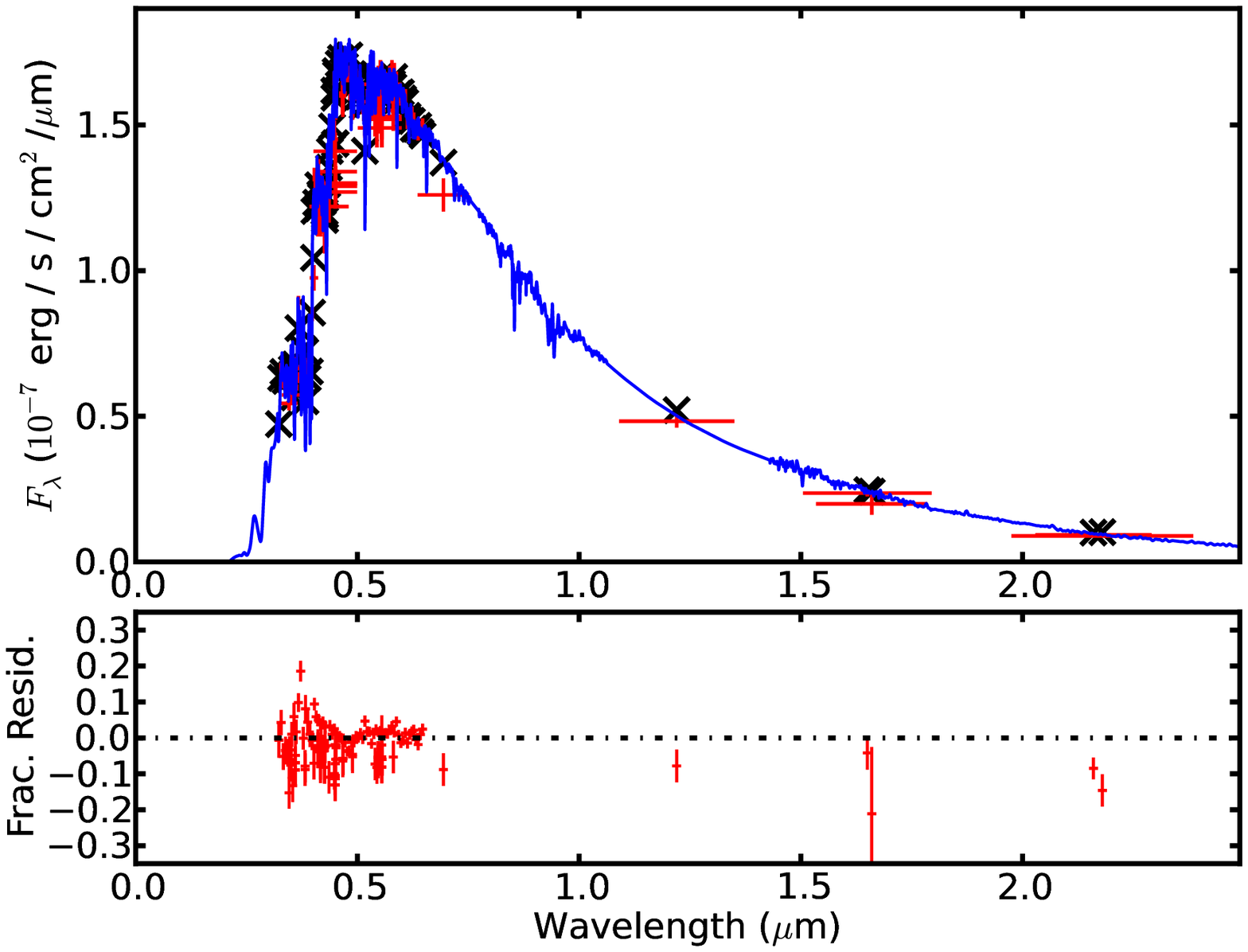} & \includegraphics[width=9.5cm]{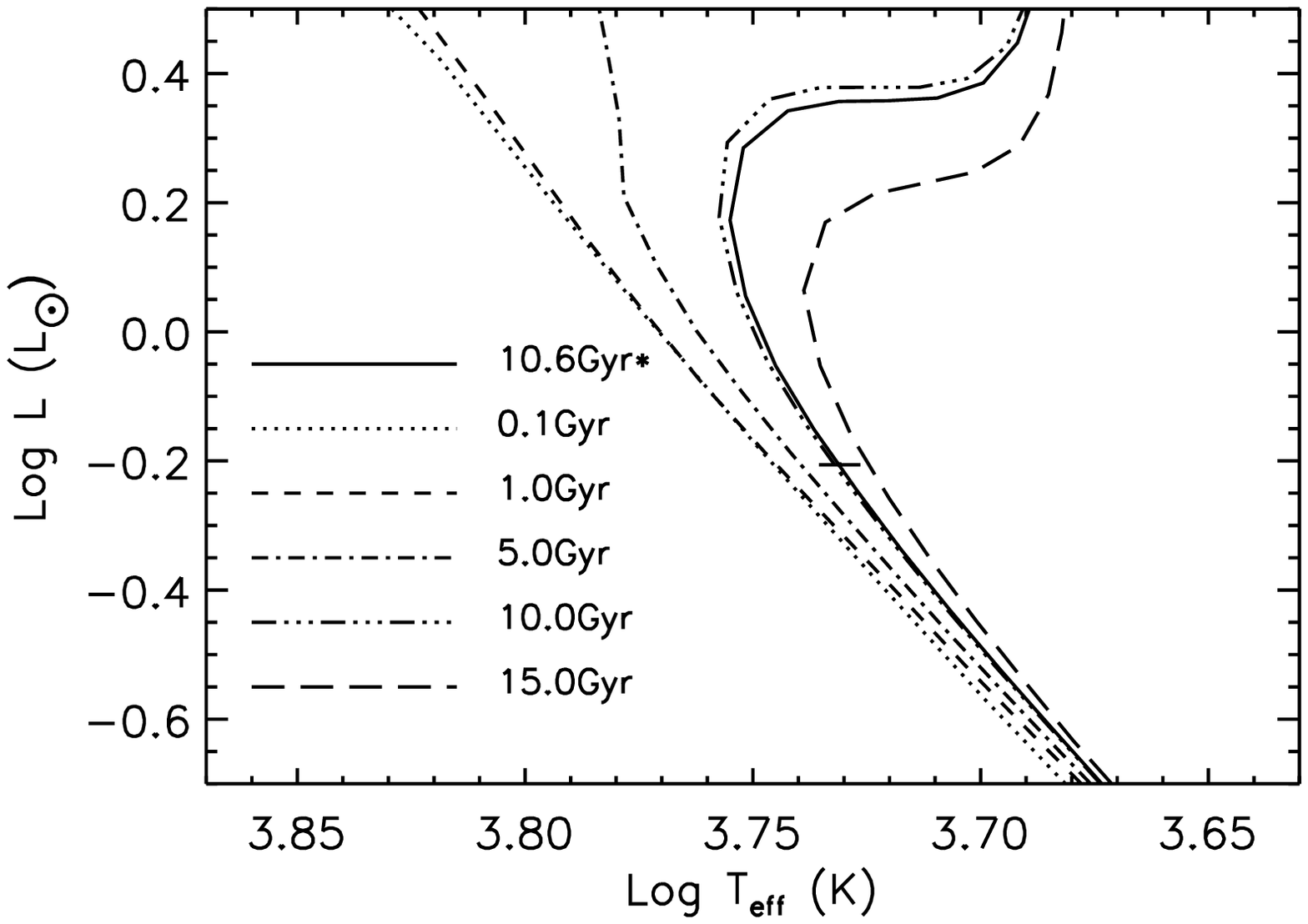}
    \end{tabular}
  \end{center}
\caption{{\it Left: -} Plot of the spectral energy distribution of HD 69830 including optical and 2MASS JHKs photometry (red +Õs, where the extent along the x-axis corresponds to filter bandwidth) fitted with stellar templates (blue line) from the library of Pickles (1998). The black x-shaped symbols represent the the value of the template at the central filter wavelength. The denser region of photometric data points around 400 nm represents the spectrophotometric data (for more detail, see Section 2.1). {\it Right -} HR diagram with Yonsei-Yale stellar evolution isochrones and a data point representing our estimated luminosity and effective temperature of HD 69830. Comparison of the measured physical properties of HD 69830 with these isochrones suggest an age of 10.6 $\pm$4 Gyr and mass of 0.863$\pm$0.043 M\sun.} 
\label{sedagefig}
\end{figure*}

We estimate the bolometric flux $F_{\rm bol}$ of HD 69830 by fitting observed optical and infrared photometry to the Pickles (1998) spectral templates (see Figure~\ref{sedagefig}). Both G8V and K0V templates from Pickles (1998) were fit to the data, with no significant difference in the $\chi^2$ value between the two. Given the preference of recent, authoritative surveys for the former spectral type (Keenan et al. 1989; Gray et al. 2006), we opted to use the values derived from the K0V fit. The template spectra were adjusted to account for overall flux level. Both narrowband and wideband photometry from 0.5 $\mu$m to 3.5 $\mu$m were used as available, including Johnson $UBV$ (Cowley 1967), $DDO$ (McClure1981; Dean 1981), Str\"{o}mgren $ubvy$ (Heck 1980; Olsen 1994), Geneva (Rufener 1976), Oja (Haggkvist 1987), IRAS (Beichman 1988), 2MASS (Cutri 2003), and Johnson $JHKL$ (Aumann et al. 1991).  Additionally, spectrophotometric data spanning 0.323 to 0.758 $\mu$m in 5 nano-meter steps (Kharitonov 1988) were included to constrain the fit. Zero-magnitude flux density calibrations were based upon the values given in Fukugita et al. (1995) and Cox et al. (2000), or the system reference papers cited above.  One minor concern regarding the photometry used herein is that it was not taken contemporaneously with our other observations. However, this concern is mitigated by noting this particular object does not exhibit significant photospheric variability (Canto Martins et al. 2011).

The resulting bolometric flux is $F_{\rm bol}$=$1.28\times10^{-7} \pm 2.57\times10^{-9} $ erg cm$^{-2}$ s$^{-1}$.  As pointed out in von Braun et al. (2014), we note that the quoted uncertainty for $F_{\rm bol}$ cannot and does not take into account unknown systematic effects caused by uncertainties in photometric magnitude zero point levels or filter profiles, correlated errors in the photometry (e.g., saturation), systematics in the spectral templates, etc. To quantify this effect and attain a more realistic representation of the uncertainty in the bolometric flux value, we follow the reasoning in sections 3.2.1$-$3.2.3 of Bohlin et al. (2014) and add in quadrature a 2\% uncertainty to our $F_{\rm bol}$ uncertainty value, which is presented in Table~\ref{props}.

To determine the effective temperature of the star we put the bolometric flux and limb darkened angular diameter into a modified Stefan-Boltzmann Law

\begin{equation}
T_{\rm eff} = 2341(F_{\rm bol}/\theta^2_{LD})^{\frac{1}{4}}
\end{equation}

\noindent where $F_{\rm bol}$ is in units of $10^{-8}$~erg cm$^{-2}$ s$^{-1}$ and $\theta_{LD}$ is in units of mas. From this, we derive an effective temperature of 5394$\pm$62 K (see Table 2).

\subsection{The Age and Mass of HD 69830}

With precise values of the luminosity, effective temperature determined from its radius and SED as well as its gravity and metallicity published in the SPOCS survey (Valenti \& Fischer 2005), we can place HD 69830 on an Hertzsprung-Russel (HR) diagram along with isochrones from the latest stellar formation models to determine the age and mass of the star. Figure~\ref{sedagefig} shows an HR diagram with a data point plotted at the measured effective temperature and luminosity of HD 69830. Also plotted are the Yonsei-Yale ($Y^2$) stellar evolution isochrones for a range of ages (0.1$-$15 Gyr, Demarque et al. 2004; Yi et al. 2001). By comparing the properties of HD 69830 to the isochrones we determine an age for the star of 10.6$\pm$4 Gyr and a mass of 0.863$\pm$0.043  M\sun (see Table~\ref{props}). The uncertainties are derived from varying the effective temperature and stellar radius by $\pm$ 1 $\sigma$ and re-estimating the age and mass. While the error bar for the age is still somewhat large, our age does rule out the youngest ages previously associated with this star (Beichman et al. 2005). 

\subsection{Determining the parameters of HD 69830 from Synthetic Spectral Fitting}

To determine the values of additional stellar properties of HD 69830 and to refine our mass and age estimates, we analyzed a set of nine archived high-resolution Keck/HIRES echelle spectra\footnote{http://www2.keck.hawaii.edu/koa/public/koa.php} collected as part of the effort to characterize the orbits of its three Neptune-mass planets. We utilized the spectral analysis tool Spectroscopy Made Easy (SME) according to the methods of Valenti \& Fischer (2005) and additional iterative tools from Valenti et al. (2009) to extract the relevant spectral properties from the nine spectra through model fits to the data. To chose which portion of the spectra to fit, we use the spectral line list of Valenti \& Fischer (2005) which includes eight segments spanning 160 \AA{} including the Mg 1b triplet region between 5165 and 5190 \AA{} and most of the region between 6000 and 6200 \AA{}. 

The model spectra were created and then fit to the data by first perturbing the effective temperature ($T_{\rm eff}$) and then the gravity ($\log g$) values. The free parameters for the first set of iterations were $T_{\rm eff}$, $\log g$, [M/H], vsin$i$, and the Fe, Si, Ni, Na, and Ti abundances. As in Valenti \& Fischer (2005) we have set the microturbulent velocity to the solar value of 0.85 km/s to minimize the errors in our [M/H] values. With each iteration of SME we also located the most appropriate spectral evolution model in the $Y^2$ grid based on the values of $T_{\rm eff}$, [Fe/H], and [$\alpha$/Fe].  The grid location probabilities were calculated assuming that the uncertainties on the parameters are normally distributed.  The model parameters returned are then the median of the distribution of masses found along with the $\pm$1 $\sigma$ limits. After an initial best fit model of each of the nine spectra was achieved, the $T_{\rm eff}$ was perturbed by $\pm$100K and two new solutions were found. The best fit models from the three effective temperature assumptions were then averaged ($\chi^2$ weighted). The $T_{\rm eff}$, [Fe/H], and [Si/H] along with V mag, distance, and bolometric correction were used to interpolate onto the $Y^2$ isochrones to retrieve the corresponding value of $\log g$. If the value of $\log g$ from this iterative method was significantly different than the value derived from the isochrone fitting, a new iteration was begun with the prior averaged parameters, but with $\log g$ fixed at the isochrone value.  The process was repeated until the gravity values converged to within 0.001 dex. 

With the permuted $\log g$ analysis, the setup was the same as the previous set of iterations except the $T_{\rm eff}$ was set and fixed to the central value from the interferometry measurements. The $\log g$ was perturbed by $\pm$0.1 dex, the procedure was run three times and the parameter values for each spectrum are the $\chi^2$ weighted average of the three models. In each iteration, a new set of parameters are found with the $\log g$ from the previous iteration and then those are used to obtain a new $\log g$ from the isochrones.  This iteration continues until the two converge ensuring a self consistent set of parameters.  The age is again determined from the $Y^2$ isochrones and the uncertainties are again derived from varying the new values of the effective temperature and stellar radius by $\pm$ 1 $\sigma$. The best fitting parameters are listed in Table~\ref{props}. As a result of the lower uncertainties on the radius and temperature of the star from SME, the age derived from spectral fitting (7.5$\pm$3) also has a smaller uncertainty but is in agreement from the age estimated from the first iteration of isochrone fitting. The effective temperature and $\log g$ estimates are consistent with values derived using other methods (i.e. $\log g$=4.40 and $T_{\rm eff}$=5402, Adibekyan et al. 2012). 

\subsection{Location of the Habitable Zone and Snow Line}

\begin{figure*}
  \begin{center}
    \begin{tabular}{c}
      \includegraphics[width=8.5cm, angle=-90]{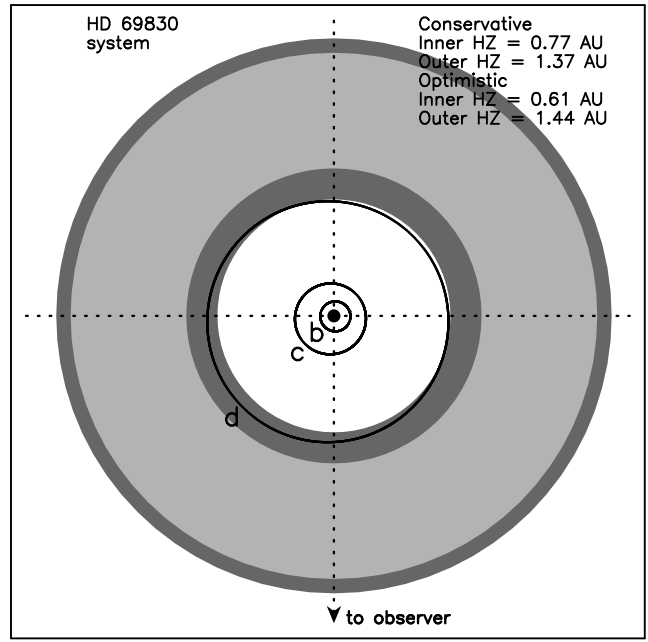}
    \end{tabular}
  \end{center}
  \caption{A top-down view of the HD 69830 system showing the extent of the habitable zone relative to the
planetary orbits (see Section 3.4). The light-gray region indicated the conservative habitable zone and the dark-gray indicated the
optimistic extension to the conservative habitable zone. The solid lines are the Keplerian orbits of the known planets. }
  \label{hzfig}
\end{figure*}

As shown by Kane (2014), the boundaries of the Habitable Zone (HZ) depend sensitively on the parameters of the host star. With this
precise value of stellar luminosity presented here, we are able to determine the location of the HD 69830 HZ with a high degree of
confidence. To do this we employ the climate models outlined in Kasting et al. (1993) and Kopparapu et al. (2013, 2014) and have been
adopted by Kane \& Gelino (2012) for the Habitable Zone Gallery\footnote{\tt http://hzgallery.org/}. These models are divided
into an $``$optimistic" and $``$pessimistic" scenario in choosing the assumed planetary climates for the determination of the inner and
outer limits of the HZ (Kane et al. 2013). The optimistic criteria assumes a $``$recent venus" and $``$early mars" for the inner and
outer boundaries while the $``$conservative" criteria assumes a $``$runaway greenhouse" and $``$maximum greenhouse" for the same boundaries. Using the conservative criteria we estimate inner and outer habitable zone boundaries of 0.767 and 1.368 AU, respectively. Using the optimistic criteria the
inner and outer habitable zone boundaries are 0.605 and 1.442 AU, respectively.  It should be noted that one of the known radial
velocity planets, HD 69830 d lies near the $``$recent venus" boundary with its orbital distance of 0.63 AU (Lovis et al. 2003, see Figure~\ref{hzfig}) and actually crosses briefly into the HZ boundary for part of its elliptical orbit. 
In addition, we use the equations described in Ida \& Lin (2005) and Kane (2011) to estimate a location for the snow line of 1.95$\pm$0.19 AU which lies outside the outer boundary of the optimistic habitable zone and well outside the location of the asteroid belt. The uncertainty is derived by error propagation using our estimated value of the mass of the star and the relationship for the location of the ice line given in Ida \& Lin (2005).

\section{Conclusions}

We present direct measurements of the diameter of HD 69830 with the CHARA array along with additional stellar parameters derived from this measurement when combined with published photometry and optical echelle spectra. They include the effective temperature, luminosity, $\log g$, metallicity and mass of the star. Three of the stellar parameters, the age, luminosity and effective temperature, were determined from two independent measurements. In all three cases the parameters are statistically consistent with each other. The values of effective temperature measured directly from the CHARA data and from the SME fits to the echelle spectra are impressively consistent with a difference of only nine degrees Kelvin (5394$\pm$62 K from the SED fit versus 5385$\pm$44 K from the spectral fit). The most notable refined stellar parameter is the age of HD 69830 which we estimated from both stellar isochrones and spectral modeling. Both of the values, 10.6$\pm$4 Gyr and 7.5$\pm$3 Gyr, respectively are consistent with each other. These values remove the possibility that this is a young star, 0.6-2 Gyr old, as previously reported by Song et al. (2000). Our age values are also in agreement with those determined through gyro-chronology and activity-age relationships (5.7-6.1 Gyr, Mamajek \& Hillenbrand 2008) which suggests that SED fits with the aid of a stellar radius from CHARA could be used to provide stellar age estimates in the absence of rotation or activity measurements. The older age for this star supports the theory that its asteroid belt cannot be primordial and must be either continuously replenished through asteroid impacts or is a short-term event (Beichman et al. 2005). Finally, our new estimate of the location of the optimistic habitable zone places it just outside the orbit of the furthest planet, HD 69830d while the snow line lies well outside the orbits of all the planets. 

With no additional planets currently known to exist at more distant orbits (C. Lovis private communication), it may take some time for RV planet search programs to be able to fully confirm the nonexistence of Jupiter-mass planets near or past the snow line. 
HD 69830 is most likely the target of new high contrast direct imaging instruments such as those being designed for the next generation ground-based extremely large telescopes or the WFIRST-AFTA program (Spergel et al. 2003), therefore, information on the presence of massive planets beyond the ice line may be on the horizon. In any case, the direct measurement of the radius of this star has proven to be an essential tool for understanding the properties of this complicated planetary system and could be applied to additional systems as they are discovered with planet search programs around bright, nearby stars such as those found with TESS (Ricker 2014). 

\begin{acknowledgements}
We thank the anonymous referee for insightful comments pertaining to this manuscript. This research has made use of the JSDC Jean-Marie Mariotti Center database, available at http://www.jmmc.fr/jsdc. The CHARA Array is funded by the National Science Foundation through NSF grants AST-0908253 and AST 1211129, and by Georgia State University through the College of Arts and Sciences. TSB acknowledges support provided through NASA grant ADAP12-0172.  This research has made use of the Habitable Zone Gallery at hzgallery.org.
\end{acknowledgements}

\begin{deluxetable}{lrrr}							
\tablecaption{Observation Logs\label{obs}}							
\tablehead{ \colhead{UT} & \colhead{Baseline} &\colhead{\# of Brackets} & \colhead{Calibrator} \\	
 Date & & & HD 	}					
\startdata							
Feb 3 2012	&	S1/W1	&	5	&	HD 66643, HD 71766	\\
Nov 3 2012	&	S1/E1	&	5	&	HD 66242, HD 71766	\\
Nov 4 2012	&	S1/E1	&	5	&	HD 66242, HD 71766	\\
\enddata							
\end{deluxetable}	

\begin{center}
\begin{deluxetable}{lcccc}					
\tablecaption{Physical properties of HD 69830\label{props}}					
\tablehead{ \colhead{Property} & \colhead{Value 1} & \colhead{Method 1$^a$} & \colhead{Value 2} & \colhead{Method 2}}		
\startdata	
Parallax (mas)	                   &	80.04$\pm$0.35	&	van Leeuwen (2007) && 	\\
$\theta_{\rm UD}$ (mas)	&	$0.655\pm0.013$	&	CHARA &&	\\
$\theta_{\rm LD}$ (mas)	&	$0.674\pm0.014$ 	&	CHARA &&	\\
Bolometric Flux (erg cm$^{-2}$ s$^{-1}$) & 1.28$\times10^{-7} \pm 2.57 \times 10^{-9} $  & SED Fit && \\
Luminosity (L\sun)	            &	0.622$\pm$0.014 L\sun	& SED Fit & 0.59$\pm$0.028  &  isochrone \\
Effective Temperature (K) &	5394$\pm$62 & CHARA+SED fit   &	5385$\pm$44 & SME\\
Age (Gyr)	&	10.6$\pm$4 &	isochrone 	&	 7.5$\pm$3  &SME	\\
Spectral Type	&	G8V	&	Gray et al. (2006) &&	\\
Radius	(R\sun) &	0.9058$\pm$0.0190 	&	CHARA &&	\\
Mass (M\sun)	&	 0.863$\pm$0.043  	& isochrone &&	\\
$[Fe/H]$	&	-0.04$\pm$0.03 	&	SME &&	\\
$[\alpha/H]$	&	0.06$\pm$0.04 	&	SME &&	\\
vsin$i$  (km/s) &  0.8$\pm$0.5 & 	SME &&	\\
$\log g$ (cgs)	&	4.49$\pm$0.06 	&	SME &&	\\
HZ boundary - pessimistic (AU) &	0.767 to 1.368	& using Kane \& Gelino (2012) 	&&\\
HZ boundary - optimistic (AU)	&	0.605 to 1.442 	& using Kane \& Gelino (2012)	&&\\
Snow line boundary (AU) &	1.95$\pm$0.19	& using Ida \& Lin (2005)	&&\\				
\enddata
\tablenotetext{a}{Because in some cases there were two independent measurements of the stellar physical parameters, we have listed them both here and have noted the method by which that parameter was determined. In all cases the multiple measurements are statistically consistent with each other.}					
\end{deluxetable}
\end{center}



\newpage

\begin{references}
\reference{} Adibekyan, V.~Z., Sousa, S.~G., Santos, N.~C., et al.\ 2012, A\&A, 545, AA32
\reference{} Aumann, H., \& Probst, R.~G. 1991 ApJ, 368, 264
\reference{} Beichman, C.~A., Neugebauer, G., Habing, H.~J., Clegg, P.~E., 
\& Chester, T.~J.\ 1988, Infrared astronomical satellite (IRAS) catalogs and atlases.~Volume 1: Explanatory supplement, 1,  
\reference{} Beichman, C.~A., Bryden, G., Gautier, T.~N., et al.\ 2005, \apj, 626, 1061
\reference{} Beichman, C.~A., Lisse, C.~M., Tanner, A.~M., et al.\ 2011, \apj, 743, 85 
\reference{} Bohlin, R.~C., Gordon, K.~D., \& Tremblay, P.-E.\ 2014, \pasp, 126, 711 
\reference{} Bonneau, D., Delfosse, X., Mourard, D., et al.\ 2011, \aap, 535, A53 
\reference{} Boyajian, T.~S., van Belle, G., \& von Braun, K.\ 2014, \aj, 147, 47 
\reference{} Boyajian, T.~S., von Braun, K., van Belle, G., et al.\ 2013, \apj, 771, 40 
\reference{} Boyajian, T.~S.\ 2009, Ph.D.~Thesis
\reference{} Canto Martins, B.~L., Das Chagas, M.~L., Alves, S., et al.\ 2011, A\&A, 530, A73
\reference{} Claret, A.\ 2000, A\&A, 363, 1081
\reference{} Cowley, A.~P., Hiltner, W.~A. \& Witt, A.~N., 1967, AJ, 72, 1334
\reference{} Cox, A.~N.\ 2000, Allen's Astrophysical Quantities,  
\reference{} Cutri, R.~M., Skrutskie,  M.~F., van Dyk, S., et al.\ 2003, ''The IRSA 2MASS All-Sky Point Source 
Catalog, NASA/IPAC Infrared Science Archive.
\reference{} Dean, J.~F.\ 1981, South African Astronomical Observatory Circular, 6, 10 
\reference{} Demarque, P., Woo, J.-H., Kim, Y.-C., \& Yi, S.~K.\ 2004, \apjs, 155, 667  
\reference{} Fukugita, M., Shimasaku, K. \& Ichikawa, T., 1995, PASP, 107, 945
\reference{} Gray, R.~O., Corbally, C.~J., Garrison, R.~F., et al.\ 2006, AJ, 132, 161
\reference{} Haggkvist, L. \& Oja, T. 1987, A\&AS, 68, 259
\reference{} Heck, A. \& Manfroid, J. 1980, A\&AS, 42, 311
\reference{} Ida, S., Lin, D.N.C. 2005, ApJ, 626, 1045
\reference{} Ji, J., Kinoshita, H., Liu, L., \& Li, G.\ 2007, \apj, 657, 1092 
\reference{} Kane, S.R. 2014, ApJ, 782, 111
\reference{} Kane, S.~R., Barclay, T., \& Gelino, D.~M.\ 2013, ApJL, 770, L20 
\reference{} Kane, S.R., Gelino, D.M. 2012, PASP, 124, 323 
\reference{} Kane, S.~R.\ 2011, Icarus, 214, 327 
\reference{} Kasting, J.F., Whitmire, D.P. \& Reynolds, R.T. 1993, Icarus 101, 108
\reference{} Keenan P.~C. \& McNeil, R.~C. 1989, ApJS, 71, 245
\reference{} Kharitonov, A.~V., Tereshchenko, V.~M., \& Knyazeva, L.~N.\ 1988, The spectrophotometric catalogue of stars.~Book of reference., by Kharitonov, A.~V.; Tereshchenko, V.~M.; Knyazeva, L.~N..~ Nauka, Alma-Ata (USSR), 1988, 478 p., ISBN 5-628-00165-1, Price 4 Rbl.~80 Kop.,  
\reference{} Kopparapu, R.~K., Ramirez, R., Kasting, J.~F., et al.\ 2013, \apj, 765, 131
\reference{} Kopparapu, R.~K., Ramirez, R.~M., SchottelKotte, J., et al.\ 2014, \apjl, 787, L29
\reference{} Lafrasse, S., Mella, G., Bonneau, D., et al.\ 2010, VizieR Online Data Catalog, 2300, 0 
\reference{} Lisse, C.~M., VanCleve, J., Adams, A.~C., et al.\ 2006, Science, 313, 635 
\reference{} Lovis, C., Mayor, M., Pepe, F., et al.\ 2006, \nat, 441, 305 
\reference{} Mamajek, E., \& Hillenbrand, L., 2008, ApJ, 687, 1264
\reference{} McClure, R.~D. \& Forrester, W.~T.  1981, PDAO, 15, 439
\reference{} Olsen, E.~H., 1994, A\&AS, 106, 257
\reference{} Pickles, A.~J.\ 1998, \pasp, 110, 863 
\reference{} Ram{\'{\i}}rez, I., Allende Prieto, C., \& Lambert, D.~L.\ 2013, \apj, 764, 78
\reference{} Ricker, G.~R.\ 2014, Search for Life Beyond the Solar System.~Exoplanets, Biosignatures \& Instruments, 3 
\reference{} Rufener, F., 1976, A\&AS, 26, 275
\reference{} Rugheimer, S., \& Haghighipour, N.\ 2007, Bulletin of the American Astronomical Society, 39, 203 
\reference{} Smith, R., Wyatt, M.~C., \& Haniff, C.~A.\ 2009, \aap, 503, 265
\reference{} Song, I., Caillault,  J.-P., Barrado y Navascu{\'e}s, D., Stauffer, J.~R., \& Randich, S.\ 2000, ApJL, 533, L41 
\reference{} Spergel, D., Gehrels, N., Breckinridge, J., et al.\ 2013, arXiv:1305.5422 
\reference{} Sturmann, J., ten Brummelaar, T.~A., Ridgway, S.~T., et al.\ 2003, \procspie, 4838, 1208 
\reference{} ten Brummelaar, T.~A., McAlister, H.~A., Ridgway, S.~T., et al.\ 2005, \apj, 628, 453
\reference{} Torres, G., Fischer, D.~A., Sozzetti, A., et al.\ 2012, \apj, 757, 161 
\reference{} Valenti, J.~A.,  Fischer, D., Marcy, G.~W., et al.\ 2009, \apj, 702, 989
\reference{} Valenti, J.~A., \& Fischer, D.~A.\ 2005, \apjs, 159, 141 
\reference{} van Leeuwen, F.\ 2007, \aap, 474, 653
\reference{} von Braun, K., Boyajian, T.~S., ten Brummelaar, T.~A., et al.\ 2011, \apj, 740, 49 
\reference{} von Braun, K., Boyajian, T.~S., Kane, S.~R., et al.\ 2012, \apj, 753, 171 
\reference{} von Braun, K., Boyajian, T.~S., van Belle, G.~T., et al.\ 2014, \mnras, 438, 2413 
\reference{} Yi, S., Demarque, P., Kim, Y.-C., et al.\ 2001, \apjs, 136, 417 
\end{references}
\end{document}